\documentclass[11pt]{article}
\usepackage[left=1in, right=1.75in]{geometry}
\usepackage[utf8]{inputenc}
\usepackage{authblk}
\usepackage{amsmath}
\usepackage{amssymb}
\usepackage{xcolor}
\usepackage[hidelinks]{hyperref} 
\title{Accelerating phase-field simulations on exascale computing systems for faster-than-real-time precipitate aging predictions
\thanks{Notice:  This manuscript has been authored in part by UT-Battelle, LLC, under contract DE-AC05-00OR22725 with the US Department of Energy (DOE). The US government retains and the publisher, by accepting the article for publication, acknowledges that the US government retains a nonexclusive, paid-up, irrevocable, worldwide license to publish or reproduce the published form of this manuscript, or allow others to do so, for US government purposes. DOE will provide public access to these results of federally sponsored research in accordance with the DOE Public Access Plan (http://energy.gov/downloads/doe-public-access-plan).}}

\author[1]{Stephen DeWitt\thanks{dewittsj@ornl.gov}}
\author[3]{David J. Gardner}
\author[2]{Philip Fackler}
\author[1]{Younggil Song}
\author[2]{Miroslav Stoyanov}
\author[3]{Carol S. Woodward}
\author[1]{Balasubramaniam Radhakrishnan}
\affil[1]{Computational Sciences and Engineering Division, Oak Ridge National Laboratory, Oak Ridge, TN 37831, USA}
\affil[2]{Computer Science and Mathematics Division, Oak Ridge National Laboratory, Oak Ridge, TN 37831, USA}
\affil[3]{Center for Applied Scientific Computing, Lawrence Livermore National Laboratory, Livermore, CA 94550, USA}

\usepackage[numbers,sort&compress]{natbib}
\usepackage{graphicx}
\usepackage{caption}
\usepackage{subcaption}
\usepackage[normalem]{ulem}
\usepackage{booktabs}
\usepackage[textwidth=1.5in]{todonotes}

\newcommand{\units}[1]{\mskip3mu\textrm{{#1}}}

\begin{document}
\date{}

\maketitle

\begin{abstract}
Three-dimensional phase-field simulations are a gold-standard for microstructure prediction for materials, but their computational cost often limits application-relevant calculations to modest domain sizes and timescales. Here, we present a holistic approach for accelerating Fourier pseudospectral phase-field simulations by combining performance-portable GPU computing, large-scale distributed-memory parallelism, and high-order implicit–explicit time integration within the MEUMAPPS C++ framework. We demonstrate this approach with a 1.5-billion-grid-point simulation of the growth and coarsening of 1,920 $\gamma''$ precipitates in a Ni–Nb–Fe alloy. The simulated seven-hour heat treatment is completed in 5.5 hours, making the calculation faster than real time and an estimated 217–501× faster than a CPU-only, first-order baseline. This performance makes three-dimensional simulations of thousands of interacting precipitates tractable, enabling quantitative studies of collective microstructural phenomena, large simulation ensembles, and real-time integration into controlling and interpreting experiments. Benchmarking shows single-node GPU speedups of up to 17.5× relative to comparable CPU resources, near-ideal strong scaling to 4096 GPUs for a 12-billion-grid-point problem, and a further 2.6–2.9× acceleration from fourth-order time integration at scientifically relevant error tolerances. These results establish a performance-portable strategy for exploiting leadership-scale GPU systems that is applicable to a broad class of Fourier pseudospectral simulations beyond phase-field models such as fluid dynamics and crystal plasticity simulations.

\end{abstract}

\section{Introduction}

Phase-field models have been demonstrated to accurately describe the dynamics of phase separation and evolution across a variety of scientific domains \cite{DeWitt2018, tonks_review, MOELANS2008268}, but 3D phase-field simulations usually require much more time to run than the underlying physics -- taking days or weeks to calculate phenomena that occur in seconds or hours \cite{POULSEN201698, GHOSH2026114323, DEWITT2017378}. 
In this article we present a multiprong approach to accelerate phase-field simualtions of solid-state precipitation in a nickel-base alloy. Combining GPU acceleration, large-scale distributed computing, and high-order implicit-explicit (ImEx) time integration, we demonstrate a faster-than-real-time simulation of aging of nearly 2000 precipitates under standard heat treatment conditions and estimate a two order-of-magnitude decrease in time-to-solution over a baseline approach. This simulation, with 1.5 billion grid points, is one of the largest phase-field simulations conducted to date.

Phase-field models provide a description of the evolution of a multiphase system governed by thermodynamic driving forces and using a thermodynamically consistent diffuse transition between phases. In metal alloys, phase-field models can describe the evolution of second-phase particles (``precipitates'') that form through solid-solid phase transformations. The evolution of precipitates can have a large influence on the properties of materials, impacting strength as well as the susceptibility to fracture and corrosion. For reviews on phase-field models both generally and for precipitation, refer to Refs. \cite{MOELANS2008268, DeWitt2018, chen_review, Shen2005, tonks_review}. 

Phase-field simulations of precipitation are computationally intensive, often requiring large physical domains to accommodate many precipitates and long time horizons. One particular area of interest is in coarsening dynamics, where the larger precipitates grow at the expense of smaller ones, and particle-resolving methods (like phase field) can be used to test the validity of mean-field models \cite{PhysRevLett.86.1259,POULSEN201698,ZHOU2014270}. 2D simulations of coarsening dynamics have shown that thousands of particles are necessary to probe how the dynamics are affected by factors such as elasticity, with Ref. \cite{PhysRevLett.86.1259} as an example where the coarsening of 4000 particles was simulated until only 100 remained. Performing these studies in 3D has proven difficult, owing to the massive computational expense. As an example, a simulation of the coarsening of 510 precipitates by Poulsen et. al \cite{POULSEN201698} required weeks to finish on 200 CPU cores.

The Fourier pseudospectral method has been widely used for phase-field simulations, especially for solid-state transformations where the ease of implementing ImEx time integration and nonlinear elasticity solvers is attractive \cite{chen_review, Shen2005, CHEN1998147, HU20011879, Durga_2013, Radhakrishnan2016, MOELANS2008268}. The Fourier pseudospectral method is a spatial discretization method where field variables are represented by a series expansion in sine and cosine functions and is part of a broader class of spectral methods \cite{boydbook}. Compared to other common spatial discretization methods for partial differential equations, such as the finite difference method and finite element methods that use a local representation of the field for derivatives, Fourier pseudospectral methods use a global basis. This global basis yields excellent (exponential) convergence of the truncation error, but means that different parallelization strategies are needed than for local methods \cite{10.1007/978-3-030-50371-0_19}. Outside of phase-field applications, Fourier pseudospectral methods are used in a wide variety of applications including: crystal plasticity, fracture, and turbulent fluid flow \cite{shanthraj_spectral, math8081385, boydbook}. Historically, pseudospectral phase-field simulations have been carried out using low-order ImEx time integration schemes. In their pioneering work, Chen and Shen \cite{CHEN1998147} developed what is now the standard ImEx splitting approach for phase-field models. They present first, second, and third order schemes where the stiff term is handled implicitly using a backward difference formula (BDF) and the nonlinear term is handled explicitly using an Adams-Bashforth (AB) scheme. In practice often only the first-order scheme is used \cite{HU20011879,PhysRevMaterials.5.053401,Radhakrishnan2016,ATTARI2023119204,provatas_elder, BOCCARDO2023112313}. 

Most large-scale phase-field simulations of precipitation have been performed on modest CPU-only resources that are small compared to modern leadership-scale supercomputers, albeit with some initial forays into larger calculations on many CPU cores or GPUs.  For example, Refs. \cite{moose_example, TEGELER2017173, POULSEN201698, prismspf} contain examples of simulations using the finite difference or finite element methods and between 16 and 1024 CPU cores. For pseudospectral methods, simulations with up to 1767 CPU cores have been reported \cite{PhysRevMaterials.5.053401}. GPU calculations using the pseudospectral method have been reported, usually with one GPU \cite{BOCCARDO2023112313, 10.1063/1.5003709, LEE2019109088, LIU2024107829}, but up to four GPUs \cite{GHOSH2024108513} or six GPUs \cite{doi:10.1177/10943420211042558}. Comparing one CPU core to one GPU, speedups of 40-58$\times$ have been observed \cite{10.1063/1.5003709, yenusah_thesis}, highlighting the potential of GPUs for these calculations. Looking outside of precipitation simulations, 100s to 10,000s of GPUs have been utilized for finite element and finite difference phase-field simulations of dendritic solidification \cite{GHOSH2022110734,Takaki_2023, 10.1145/2063384.2063388} and  65,536 CPU cores have been utilized for pseudospectral phase-field crystal simulations \cite{Pinomaa_2024}. These examples highlight the untapped opportunity for pseudospectral phase-field simulations to leverage the 10,000s of GPUs available on modern supercomputers.

In this article we present a new holistic approach for accelerating phase-field simulations of precipitation using the Fourier pseudospectral method. Although this work focuses on phase-field simulations, the approach is broadly applicable to other applications of the Fourier pseudospectral method. The contributions of this work are five-fold:
\begin{enumerate}
    \item A demonstration of the decrease in calculation time using GPUs compared to only CPUs, with up to a 17.5$\times$ speedup on 8 GPUs versus 64 CPU cores.
    \item Strong and weak scaling tests from 8 to 8192 GPUs, demonstrating near-ideal strong scaling up to 4,096 GPUs for a simulation with 12 billion grid points.
    \item An investigation of the use of first- through fourth-order implicit-explicit additive Runge--Kutta (ImEx-ARK) time integration schemes, demonstrating a 2.9$\times$ speed-up for the fourth-order method over the standard first-order method for science-relevant levels of error.
    \item A demonstration of the above methods to accelerate a phase-field simulation of precipitation by 217-501$\times$, making the simulation faster than the corresponding real-time experiment.
    \item A discussion of the implementation of the above approaches in the MEUMAPPS C++ software framework.
\end{enumerate}

\section{Results and Discussion} \label{sec:results}

\begin{figure}
     \centering
     \includegraphics[width=0.8\textwidth]{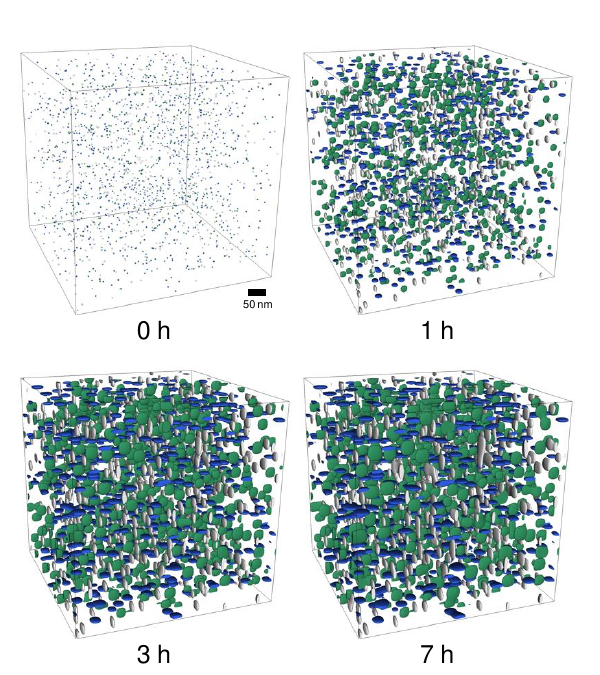}
     \caption{Snapshots of the evolution of $\gamma''$ precipitates in a Ni-Fe-Nb alloys over the course of seven simulated hours of aging at 700 C. The color indicates each of the three orientation variants of $\gamma''$ found in this system. At the start of the simulation 1920 nuclei are present split between the orientation variants. These nuclei grow for approximately the first 3 h of aging and then the system transitions into coarsening.}
     \label{fig:demo}
\end{figure}

To demonstrate the computational performance of our holistic approach to accelerating Fourier pseudospectral phase-field simulations, we present a simulation of the growth and coarsening of nearly 2000 precipitates on 1024 GPUs. We choose a technologically relevant setting for this demonstration -- the evolution of $\gamma''$ precipitates in a Ni-Fe-Nb alloy. This alloy has been used as a ternary surrogate for IN718 \cite{PhysRevMaterials.5.053401, osti_1883850}, a common precipitation-hardened nickel-base superalloy used for applications where high strength is needed at high temperatures, and $\gamma''$ precipitates are the most important strengthening phase for IN718 \cite{ZHOU2014270}. Due to its industrial importance, the evolution of $\gamma''$ in binary and ternary surrogates of IN718 has been the subject of a number of studies using phase-field simulations \cite{PhysRevMaterials.5.053401, osti_1883850, ZHOU2014270, SCHLEIFER2020106745, LASKOWSKI2021158630, SHI2019220, Ji20163235}. The model parameters we use for this material system are taken from \cite{osti_1883850}. All simulations are carried out at 700 C, a typical heat treatment temperature for IN718 for precipitation hardening. We simulate 7h of evolution of 1920 $\gamma''$ precipitates distributed over the phase's three orientation variants in a 576 nm $\times$ 576 nm $\times$ 576 nm domain (a 1152$^3$ simulation grid). Snapshots of the simulated evolution are given in  Fig.~\ref{fig:demo}. Starting with small nuclei in a supersaturated matrix at $t=0$, the system is in a growth mode until approximately $t=3$ h, when the volume fraction of precipitates stabilizes. From $t=3$ h to the end of the simulation ($t=7$ h), the system undergoes coarsening, with some precipitates growing while others dissolve. 

With over 1.5 billion grid points, this simulation is among the largest phase-field simulations to date \cite{plotkowski_imr, GHOSH2026114323, Takaki_2023} and, with the approaches described in this article, was completed in less than 5.5 hours of wall time on 1024 GPUs (128 nodes on the Frontier supercomputer at the Oak Ridge Leadership Computing Facility). This simulation runs faster than the associated experiment, which opens up the potential for performing first-of-a-kind experiments (e.g., at x-ray or neutron sources) with simulations running alongside the experiment to assist in interpreting results and/or to steer the experiment.

Using the analysis in the following sections, we estimate a transformational decrease in the time-to-solution from months to hours in the large-scale growth and coarsening simulation in Fig. \ref{fig:demo} from combining GPU acceleration (Sec.~\ref{sec:single_node_performance}) with high-order time integration (Sec.~\ref{sec:time_integration}) and scaling to a large number of nodes (Sec.~\ref{sec:scaling}) compared to a baseline scenario using a first-order time integration scheme and limited, CPU-only computing hardware. The methods that underpin these three acceleration approaches are discussed in Sec. \ref{sec:methods}.

We estimate the overall speedup using multiplicative factors for each of the three acceleration approaches. For GPU acceleration, we estimate a speedup of 7.6$\times$ to 17.5$\times$ based on the single node CPU-GPU comparisons on Frontier in Sec.~\ref{sec:single_node_performance}. For high-order time integration, we estimate a speedup of 2.6$\times$ relative to the first-order scheme with $\Delta t=0.125$s (scaling the 2.9$\times$ value from Sec.~\ref{sec:time_integration} given a decrease in $\Delta t$ from $2.25$s to $2.0$s with the high-order method). 
Combining these two factors, we expect a CPU configuration using 128 Frontier nodes (8192 CPU cores) with the first-order scheme to be 19.8$\times$ to 45.5$\times$ slower (4.5 to 11 \emph{days} to complete). As discussed in the introduction, 8192 CPU cores is far more than have typically been used for phase-field simulations of precipitation. Therefore, we choose 512 CPU cores (eight Frontier nodes) as our baseline level, which is within the common range for phase-field simulations of precipitation in the literature. The baseline choice is also motivated by the memory requirements of simulations on a 1152 $\times$ 1152 $\times$ 1152 grid, running on substantially fewer CPU cores may not be possible. 
In this more limited setting with eight Frontier nodes, we estimate that multi-node scaling provides an additional 11$\times$ speedup (equal to the speedup observed in Sec.~\ref{sec:scaling} going from 8 to 128 nodes). Therefore, multiplying the three parts of the estimated speedup yields a 217$\times$ to 501$\times$ speedup for the demonstration simulation configuration over the baseline scenario. As such, a simulation like the one presented in Fig.~\ref{fig:demo} would take between 50 and 115 \emph{days} rather than 5.5 hours using the methods presented in this article.

In the following sections we discuss the three components of our acceleration approach in detail and provide performance benchmarks for each component that yield the per-approach acceleration factors used in the preceding analysis.

\subsection{Improved performance via GPU acceleration} \label{sec:single_node_performance}

The first of the three components of our acceleration strategy is improved performance through GPU acceleration. Our GPU acceleration strategy includes acceleration of three aspects of the problem: the use of performance portable data structures and parallel execution patterns (described in Sec. \ref{sec:kokkos}), GPU-accelerated FFTs (described in Sec. \ref{sec:heffte}), and GPU-accelerated time integration (described in Sec. \ref{sec:sundials}). Using these approaches the data remains resident on the GPU throughout the simulation, eliminating the costly transfers between the GPU and CPU that bottleneck performance. Aside from some one-off processes during initialization, the only time field data is transferred from the GPU to the CPU is when writing outputs to file for visualization.

\begin{figure}
     \centering
     \includegraphics[width=0.6\textwidth]{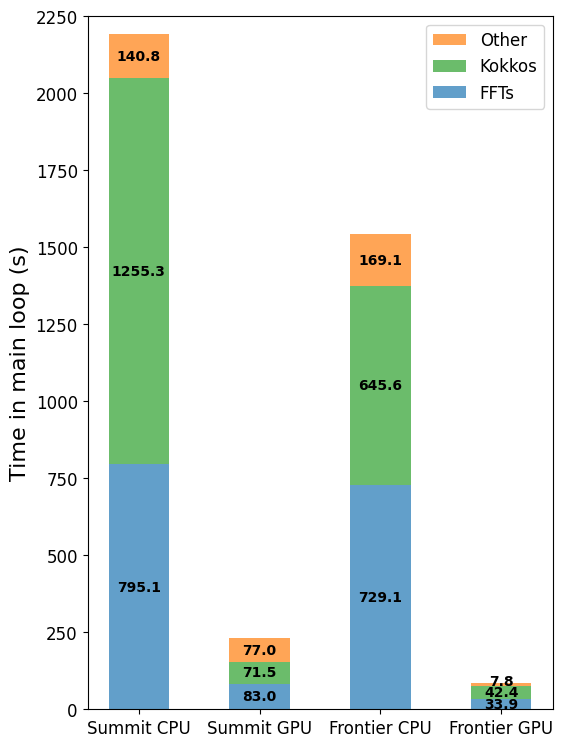}
     \caption{Single-node profiling results for the time spent in FFT library calls, Kokkos loops, and other operations for CPU-only and GPU calculations on the OLCF Summit and Frontier supercomputers.}
     \label{fig:performance-detail}
\end{figure}

To demonstrate the performance increase for this fully GPU-resident approach, we perform single-node benchmark simulations using CPU-only and GPU-accelerated simulations on the OLCF Summit and Frontier supercomputers (refer to Sec. \ref{sec:computing_systems} for details of the computing systems). The key results of this investigation are presented in Tables~\ref{tab:summit-performance} and \ref{tab:frontier-performance} and Fig.~\ref{fig:performance-detail}. Here all performance tests use the full compute capabilities of the node, comparing all of the CPU cores to all of the GPUs. This type of full-node comparison gives performance information relevant to the practical speedup of using a GPU-accelerated node versus a CPU-only node. 

The benchmark problem for this section is a version of the $\gamma''$ precipitation simulation presented above, but for three smaller grid sizes: 168$^3$ (a 84 nm cube), 336$^3$ (a 168 nm cube), and 504$^3$ (a 252 nm cube). Multiples of 168 grid points per direction were chosen to permit efficient MPI domain decompositions over 6, 8, 42, and 64 MPI tasks (the number of full-node MPI tasks for Summit GPUs, Frontier GPUs, Summit CPUs, and Frontier CPUs, respectively). The $336^3$ domain is the largest domain that will fit in Summit GPU memory for a single node, and the $504^3$ domain is the largest domain that will fit in Frontier GPU memory for a single node. On Summit, the CPU calculations range from 112,896 to 3,048,192 points per MPI rank and the GPU calculations range from 790,272 to 21,337,344 points per MPI rank. On Frontier, the CPU calculations range from 74,088 to 2,000,376 points per MPI rank and the GPU calculations range from 592,704 to 16,003,008 points per MPI rank. Benchmark simulations in all three domain sizes had a consistent number density of precipitates (6 in the 168$^3$ domain, 12 in the 336$^3$ domain, and 18 in the 504$^3$ domain), with pseudorandom positions and pseudorandom orientation variant assignment. Each benchmark test is conducted for 120 simulated seconds, using third-order ImEx time integration (see Sec. \ref{sec:time_integration}) and a fixed time step size of 1.25 s. Given the fixed number of grid points, fixed time step size, and consistent number of iterations for the nonlinear mechanics solver, preliminary tests indicated that 120 second simulations had per-time-step timings representative of longer simulations such as the simulation in Fig. \ref{fig:demo}.

\begin{table}[h]
    \centering
    \fontsize{9pt}{9pt}\selectfont
    \begin{tabular}{cc|cccc}
        \hline
        \textbf{Node type} & \textbf{Domain} & \textbf{Total (s)} & \textbf{Per point (s)} & \textbf{MPI (s, \%)} & \textbf{GPU Speedup} \\
        \hline
        CPU & 168$^3$ & 261  & $5.50 \times 10^{-5}$ & 19.8  (7.6\%) & - \\
        GPU & 168$^3$ &  55.4  & $1.17 \times 10^{-5}$ & 11.3 (20\%) & 4.7$\times$  \\
        \hline
        CPU & 336$^3$ & 2195  & $5.79 \times 10^{-5}$ & 133  (6.0\%) & - \\
        GPU & 336$^3$ & 236  &  $6.22 \times 10^{-6}$& 66.4 (28\%) & 9.3$\times$ \\
        \hline
        CPU & 504$^3$ & 5108  & $3.99 \times 10^{-5}$ & 284 (5.6\%)  & -  \\
        GPU & 504$^3$ & -        &  - & -        & -  \\
        \hline
    \end{tabular}
    \caption{Single-node CPU and GPU timings on Summit (44-core IBM Power9 CPU, NVIDIA V100 GPUs). No GPU performance data is available for the $504^3$ domain due to the memory limitations of the V100 GPUs.}
    \label{tab:summit-performance}
\end{table}

\begin{table}[h]
    \centering
    \fontsize{9pt}{9pt}\selectfont
    \begin{tabular}{cc|cccc}
        \hline
        \textbf{Node type} & \textbf{Domain} & \textbf{Total (s)} & \textbf{Per point (s)} & \textbf{MPI (s, \%)} & \textbf{GPU Speedup} \\
        \hline
        CPU & 168$^3$ & 147  & $3.09 \times 10^{-5}$ & 20.7  (14\%) & - \\
        GPU & 168$^3$ &  19.2  & $4.04 \times 10^{-6}$ & 2.65 (14\%) & 7.6$\times$  \\
        \hline
        CPU & 336$^3$ & 1548  & $4.08 \times 10^{-5}$ & 247 (16\%) & - \\
        GPU & 336$^3$ & 88.3  &  $2.33 \times 10^{-6}$& 23.4 (27\%) & 17.5$\times$ \\
        \hline
        CPU & 504$^3$ & 5117  & $4.00 \times 10^{-5}$ & 641 (13\%)  & -  \\
        GPU & 504$^3$ & 302        &  $2.36 \times 10^{-6}$ & 79.0 (26\%)        & 16.9$\times$   \\
        \hline
    \end{tabular}
    \caption{Single-node CPU and GPU performance on Frontier (64-core AMD EPYC CPU, AMD MI250X GPUs).}
    \label{tab:frontier-performance}
\end{table}

Table \ref{tab:summit-performance} shows the single-node performance on the Summit supercomputer, where each node has 42 usable IBM Power9 CPU cores (with 2 cores reserved for system processes) and 6 NVIDIA V100 GPUs. On the CPUs, the total wall time increases as the domain size increases, as expected. The average time-per-grid-point is approximately flat at 55.0-57.9 ms for the two smaller domain sizes and then decreases slightly to 39.9 ms for the largest domain size. This decrease may be due to the increased work per rank in the compute-heavy portions of the calculation hiding overheads elsewhere. On the GPUs, the total wall time (and accordingly the average per-point times) are much faster than on CPUs, yielding a 4.7$\times$ speedup for the 168$^3$ domain and a 9.3$\times$ speedup for the 336$^3$ domain. The time per-grid-point decreases substantially from the 168$^3$ domain to the 336$^3$ domain. This observation is consistent with previous findings that GPU performance degrades with low thread occupancy. The fraction of total wall time spent on MPI communication decreases slightly with increasing system size on CPU, likely due to a more favorable ratio of volume (work) to surface area (communication) scaling. On CPUs MPI communication costs are low, always under 8\% of total wall time. The total time spent on MPI communication for the GPU calculations is approximately half of the time spent for the CPU calculations. Possible reasons for communication being faster on GPUs could be the faster memory speeds on the GPUs and/or the reduced number of messages (6 MPI ranks on GPUs versus 42 ranks on CPUs). The decrease in MPI time is less than the overall GPU speedup, so the fraction of time spent on communication increases relative to the CPU-only tests. Different from the CPU-only tests, the fraction of time spent on MPI communication increases from the 168$^3$ domain to the 336$^3$ domain. Overall, these tests demonstrate a significant advantage to using the GPUs on Summit for an appropriately sized problem, with more than a 9$\times$ speedup using all 6 of the V100 GPUs on the node compared to solely the 42 CPU cores.

Table \ref{tab:frontier-performance} shows the single-node performance on the Frontier supercomputer, where each node has 64 NVIDIA EPYC CPU cores and 8 AMD MI250X GPUs. The larger memory of the MI250X compared to the V100 allows all three domain sizes to be tested on the Frontier GPUs. On the CPUs, the calculations on Frontier were slightly faster than those on Summit, requiring 30.9-40.8 ms per grid point. On the GPUs, the per-grid-point time is 4.04 ms for the 168$^3$ domain, and it decreases to about 2.3 ms for the larger domains. This same trend is observed on Summit where the per-grid-point time improves with higher thread occupancy. The per-grid-point performance then saturates at the 336$^3$ domain, suggesting that the GPU is at full occupancy. The GPU speedup on Frontier is even larger than observed for Summit -- 7.6$\times$ for the smallest domain and about 17$\times$ for the larger domains. In absolute terms, the time spent on MPI communication is similar or higher for CPU calculations on Frontier compared to Summit and lower for GPU calculations on Frontier compared to Summit. Frontier's improved MPI communication on GPUs compared to Summit may be due to a difference in the node architectures; Frontier has direct connections between its GPUs, while Summit has direct connections between groups of three GPUs which are in turn connected via the CPU. The reduction in MPI communication time is key to the improved GPU speedup on Frontier compared to Summit. For example, if the Frontier GPU calculation for the 336$^3$ domain had the same 66.4 s MPI time as the Summit calculation, the GPU speedup would only be 11.8$\times$ rather than 17.5$\times$. Overall, these tests demonstrate an even larger benefit to using the GPUs on Frontier than on Summit for an appropriately sized problem, with more than a 17$\times$ speedup using all 8 of the MI250X GPUs on the node compared to solely the 64 CPU cores.

Comparing the total wall time for GPU calculations on Summit and Frontier, the Frontier calculations were 2.9$\times$ faster and 2.7$\times$ faster than the Summit calculations for the 168$^3$ and 336$^3$ domains, respectively.
The improved performance on the Frontier nodes is expected due to the higher nominal peak performance of the Frontier nodes (191.2 double-precision TFLOPS) versus the Summit nodes (46.8 double-precision TFLOPS) and the higher nominal memory bandwidth on Frontier (1.6 TB/s) versus Summit (900 GB/s). While the observed speedup of 2.7-2.9$\times$ is less than the 4.1$\times$ increase in nominal peak FLOPS, it is larger than the 1.8$\times$ increase in memory bandwidth. MEUMAPPS C++ performance is in line with previous comparisons of Summit and Frontier showing a speedup of 1.7-6$\times$ for applications co-developed with supercomputer facility staff \cite{10.1007/978-3-031-32041-5_10}.

Figure \ref{fig:performance-detail} gives a breakdown of the timings for groups of operations in the main loop for the 336$^3$ domain calculations. Profiling data identifies the time spent on fast Fourier transform (FFT) calls to the heFFTe library and on loops using the Kokkos library. Together these two groups of operations reflect 65-93\% of the total wall time of the main loop, with roughly an equal split between them. GPU speedups are observed on both Summit and Frontier nodes for FFT calls, Kokkos loops, and for the remaining (``other'') operations. On the Summit node, FFT calls have a 9.6$\times$ speedup, Kokkos loops have a 17.6$\times$ speedup, and the other operations have a 1.8$\times$ speedup. On the Frontier node, FFT calls have a 21.5$\times$ speedup, Kokkos loops have a 15.2$\times$ speedup, and the other operations have a 14.3$\times$ speedup. These results demonstrate that the overall GPU speedup observed in Tables \ref{tab:summit-performance} and \ref{tab:frontier-performance} is achieved by GPU speedups in both the FFTs and the Kokkos loops. 

The only case where substantial GPU speedups were not observed was for the ``other'' category on the Summit node. This poor speedup of 1.8$\times$ was observed consistently during testing and was not a spurious result from a single test. Of the 77 s spent in this category, the majority of the time has no specific entry associated with it (e.g., output, Kokkos deep copies) and region profiles suggest that this time is distributed across multiple functions. With the much larger GPU speedups for the FFT and Kokkos loops, the time spent on other operations increases from 6\% to 35\% of the time in the main loop, causing a meaningful impact on overall performance. Identification of the specific operations driving poor GPU speedup of the ``other'' category is left for future work. The remaining test cases in this article focus on the newer, higher performing Frontier supercomputer, where the ``other'' category exhibits similar GPU speedups to FFTs and Kokkos loops.

In summary, the single node tests demonstrate that the performance portability strategies in MEUMAPPS C++ are effective in generating large GPU speedups (4.7-17.5$\times$) on both NVIDIA and AMD GPUs. As expected, the GPU speedup improves as the domain size, and therefore GPU occupancy, increases. The observed GPU speedup is due to GPU speedups of both FFT calls to the heFFTe library and for loops using the Kokkos library.

\subsection{Improved performance via HPC scaling} \label{sec:scaling}
Now we turn from single-node performance to scaling across multiple nodes, the second component of our acceleration strategy. Scaling to multiple nodes requires distributed memory parallelism, with explicit communication between each computational node. Minimizing the time the simulation spends waiting for data from another node is the driver for efficient multi-node scaling. For the Fourier pseudospectral method that we employ, nearly all of the distributed memory communication occurs during FFT operations, and therefore our interface to a scalable FFT library (heFFTe, as described in Sec. \ref{sec:heffte}) is the key enabler for reducing simulation times with multi-node simulations.

\begin{figure}
     \centering
     \includegraphics[width=0.6\textwidth]{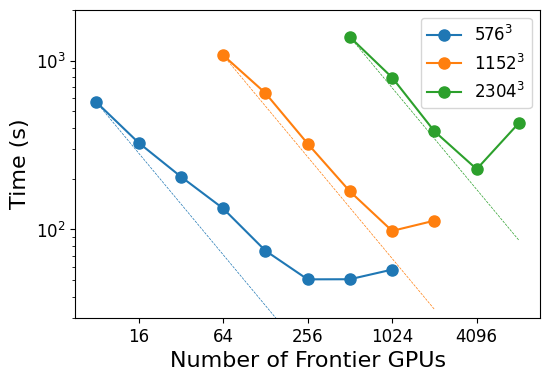}
     \caption{Combined strong and weak scaling plot for MEUMAPPS C++ on the OLCF Frontier supercomputer.}
     \label{fig:scaling}
\end{figure}

The benchmark problem setting is the same as for the single-node tests, the growth of pseudorandomly placed $\gamma''$ particles with fixed initial number density for multiple domain sizes. We focus our benchmarking on the newer Frontier supercomputer, with scaling behavior presented in Fig.~\ref{fig:scaling}. Three domain sizes are considered, testing up to 12.2 billion grid points on 8192 GPUs. The smallest domain, 576$^3$, is close to the limit of what will fit into GPU memory on one Frontier node. The other two sizes, 1152$^3$ and 2304$^3$, are factors of two and four larger in each dimension (8 and 64 times more total grid points). A combined strong and weak scaling test \cite{GHOSH2022110734} was performed with these three domain sizes, and the results are shown in Fig.~\ref{fig:scaling}. In this combined test, the lowest number of GPUs for each domain size is chosen to follow a weak scaling test (where the number of GPUs increases proportionally with the total number of grid points), and then a strong scaling test (increasing GPUs counts for a fixed total number of grid points) is performed for each domain size, with successively doubling of the GPU count until the time to solution increases. With this test matrix, corresponding places along the strong scaling tests for each domain size constitute weak scaling tests (i.e., moving left to right in Fig.~\ref{fig:scaling}, the first data point from each domain size form a weak-scaling test with the highest occupancy, the second data point from each domain size form a weak-scaling test with the second highest occupancy, etc.). Tests on fewer GPUs for the 1152$^3$ and 2304$^3$ domain sizes are not possible due to the GPU memory limitations.

Starting with the strong scaling results, Fig.~\ref{fig:scaling} demonstrates efficient strong-scaling to thousands of GPUs. For the 576$^3$ domain, the test starts on 8 GPUs (one Frontier node) and maintains a parallel efficiency above 70\% up through 32 GPUs. From 64 through 256 GPUs the scaling is sub-ideal (35-53\% parallel efficiency), but the time-to-solution continues to steadily decrease. At 512 GPUs the time-to-solution is barely faster than at 256 GPUs, and at 1024 GPUs overhead from distributing the computational finally increases enough that the time-to-solution increases. For the 1152$^3$ domain, the test starts on 64 GPUs and parallel efficiency above 70\% is maintained up through 1024 GPUs. At 2048 GPUs, the scaling abruptly worsens and the time-to-solution increases. For the 2304$^3$ domain, the test starts on 512 GPUs and parallel efficiency above 70\% is maintained up through 4096 GPUs. At 8192 GPUs, the scaling abruptly worsens and the time-to-solution increases. 

For the weak scaling results, Fig.~\ref{fig:scaling} shows consistent deviation from ideal scaling. The relevant weak scaling tests are the first four with the highest occupancy, where strong scaling is near-ideal for all three domain sizes. For the weak scaling test at the highest occupancy (23,887,872 grid points per GPU), the time to solution increases from 568s to 1077s to 1379s for the 576$^3$, 1152$^3$, and 2304$^3$ domain sizes, respectively, corresponding to parallel efficiencies of 53\% to 41\%. At the second highest occupancy (11,943,936 grid points per GPU), the time to solution increases from 326 s to 643 s to 793 s (parallel efficiencies of 51\% and 41\%). At the third highest occupancy (5,971,968 grid points per GPU), the time to solution increases from 206 s to 322 s to 381 s (parallel efficiencies of 64\% and 54\%). At the fourth highest occupancy (2,985,984 grid points per GPU), the time to solution increases from 133 s to 168 s to 228 s (parallel efficiencies of 79\% and 59\%).

The scaling tests indicate strengths and limits of the MEUMAPPS C++ framework for use on GPU supercomputers. For a large enough problem size (2304$^3$ in our tests), MEUMAPPS C++ can efficiently use at least 4096 GPUs. The per-GPU problem size where strong-scaling worsens is in the range of 1,492,992 to 2,985,984 grid points per GPU. Generally, a 6-11$\times$ reduction in the time to solution was attained by adding GPUs beyond the minimum required for the problem to fit in GPU memory. The parallel performance for weak scaling is in the 40-80\% range. The 3D FFT algorithm is known to have non-ideal weak scaling, going as $\mathcal{O}(N^3 \log(N))$ \cite{doi:10.1137/11082748X}. However, for the problem sizes examined here, the theoretical non-ideality from the $\log(N)$ would only reduce the parallel efficiency to $\sim$80\%. Benchmarks from the heFFTe library on the Summit supercomputer are consistent with those theoretical expectations with 80\% parallel efficiency for 11.2 million grid points per GPU \cite{10.1007/978-3-030-50371-0_19}. The decreased weak scaling performance in MEUMAPPS C++ may be due to increased GPU register pressure resulting from the many fields being stored in GPU memory simultaneously and is an area for future optimization within the code.

\subsection{Improved performance via time integration approach} \label{sec:time_integration}
Finally, we turn to the third component of our acceleration strategy, high-order ImEx time integration. The computational cost of a Fourier pseudospectral method is approximately proportional to the number of right-hand side (RHS) evaluations, with each time step during the simulation involving one or more RHS evaluations depending on the time integration method and order. The total number of RHS evaluations can be decreased either by reducing the number of time steps (i.e. taking larger time steps) or reducing the number RHS evaluations per time step (i.e. lowering the order of the method). Using additive Runge-Kutta methods (described in Sec. \ref{sec:sundials} and implemented in the SUNDIALS library), we find that for physically relevant error levels that the simulations are faster with large, high-order time steps than the common \cite{HU20011879,PhysRevMaterials.5.053401,Radhakrishnan2016,ATTARI2023119204,provatas_elder, BOCCARDO2023112313} approach of small, first-order time steps.

To demonstrate the performance differences across time-step size and method order, we perform single-node benchmark simulations using GPU-accelerated nodes on the Frontier supercomputer. The benchmark problem for this section involves two growing precipitates for a single $\gamma''$ orientation variant, the smallest, simplest scenario of precipitate evolution that is more complex than monotonic growth of a single precipitate. This test is performed in a cubic computational domain with 32 nm in each dimension and a 0.5 nm grid spacing (i.e., 64$^3$ domain). One precipitate seed with a 2.5 nm radius is placed at (10 nm, 15.5 nm, 15.5 nm) and a second seed with a 3 nm radius is placed at (20.5 nm, 20.5 nm, 20.5 nm). 1500 s of heat treatment was simulated, enough to see the particles complete the growth phase of evolution (where the precipitates near the equilibrium volume fraction) and begin coarsening (where the larger particle grows at the expense of the smaller particle). The small size and short length of this test case makes reference simulations with a very small time step computationally tractable. 

\begin{figure}
     \centering
     \begin{subfigure}[b]{0.6\textwidth}
         \centering
         \includegraphics[width=\textwidth]{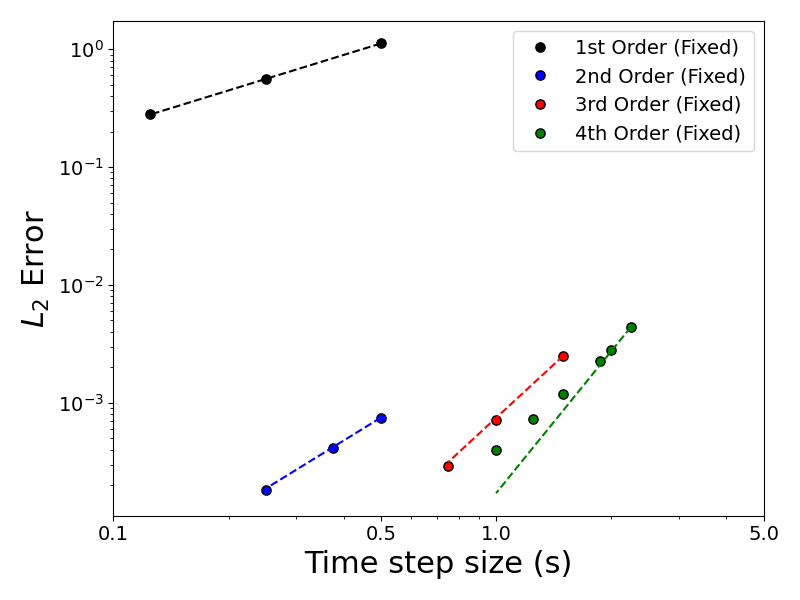}
         \caption{Error convergence}
         \label{fig:time_integration_convergence}
     \end{subfigure}
     \begin{subfigure}[b]{0.6\textwidth}
         \centering
         \includegraphics[width=\textwidth]{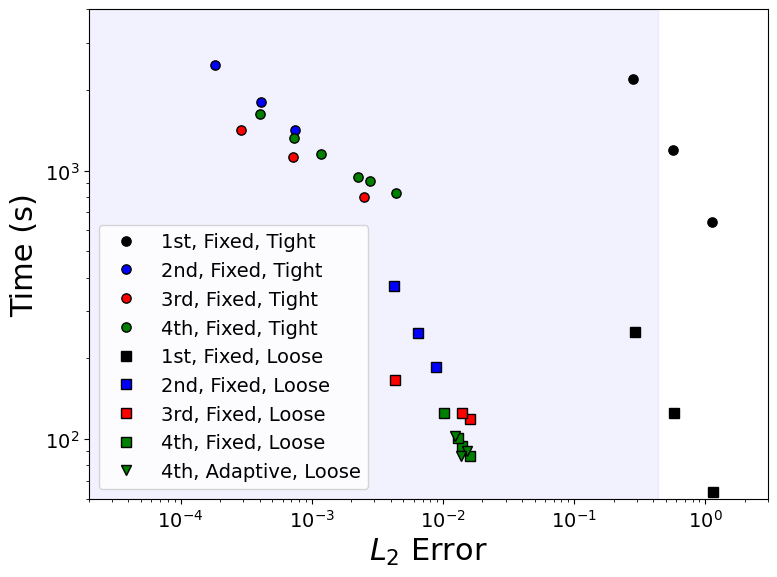}
         \caption{Method efficiency}
         \label{fig:time_integration_performance}
         \end{subfigure}
        \caption{Results evaluating different time integration methods/configurations with the two-particle test case. a) A plot of the $L_2$ error for various fixed time step sizes using tight elasticity tolerances. The markers denote the results of the numerical experiments and the dashed lines denote the expected reduction of error given the order of accuracy of the method. b) A plot of the time-to-solution versus the $L_2$ error for different time integrator orders with fixed or adaptive step sizes and a loose or tight elasticity solver tolerance. The blue region denotes simulation results with $L_2$ error below the physically relevant level.}
        \label{fig:time_integration}
\end{figure}

Using the two-particle test case we examine the error convergence and computational performance of our time integration strategy using methods ranging from first order to fourth order. The key results of this investigation are presented in Fig.~\ref{fig:time_integration}. 

Figure \ref{fig:time_integration_convergence} shows the convergence of the $L_2$ error of the simulations with decreasing time step size ($\Delta t$). The error is calculated as the $L_2$ norm of the difference between the solution and a reference solution at the end of the simulation ($t=1500\units{s}$). The reference solution is from a highly resolved simulation using the third order accurate method with $\Delta t=0.25\units{s}$. All simulations in Fig.~\ref{fig:time_integration_convergence} were carried out with a tight elasticity solver tolerance ($\epsilon_{el} = 10^{-16}$) such that the differences between the test results and the reference solution is dominated by temporal discretization error. For the first, second, and third order methods the expected order of accuracy is clearly observed. With the fourth order method, we observe fourth order convergence between the three largest time steps considered ($\Delta t = 1.875\units{s}, 2.0\units{s},2.25\units{s}$). However, this trend does not hold for the remaining three time step sizes ($\Delta t = 1.0\units{s}, 1.25\units{s}, 1.5\units{s}$) where the observed order of accuracy falls between second and third order. A degradation in the observed order of accuracy could in principle be caused by an inadequate reference solution, where the difference between test solutions and the reference is dominated by temporal discretization error in the reference rather than the test solution, but is unlikely to be the case here. Assuming the observed third order convergence continues to hold for the nominally third order method, extrapolating from the error at $\Delta t = 1.875\units{s}$ the expected $L_2$ error for the reference solution is $\sim10^{-5}$, 40$\times$ smaller than the measured error for the smallest time step size for the nominally fourth-order method and 120$\times$ larger than the measured error for the time step size where the nominally fourth-order method deviates from the fourth-order trend line. Instead, we believe the degradation in the observed convergence rate is due to order reduction when integrating stiff systems \cite{prothero1974stability, sanz1986convergence, verwer1986convergence}.

In addition to the convergence tests, as a secondary check, we manually implemented the first-order ImEx Euler method, utilized by other phase-field models \cite{CHEN1998147, HU20011879,PhysRevMaterials.5.053401,Radhakrishnan2016,ATTARI2023119204,provatas_elder, BOCCARDO2023112313}, into MEUMAPPS C++ and compared the results to those obtained using the SUNDIALS library implementation of the same method (i.e., those shown in Fig.~\ref{fig:time_integration_convergence}). The pointwise absolute error between the two solutions is close to floating-point precision, less than $10^{-16}$. Together, the convergence tests and the pointwise error verification provide strong evidence that the high-order time integration methods proposed here behave as expected and are correctly implemented in MEUMAPPS C++.

The approximate maximum time step sizes reported in Fig.~\ref{fig:time_integration_convergence} vary significantly with the order of accuracy of the integration method and were found to be $0.5\units{s}$ for first and second order methods, $1.5\units{s}$ for the third order method, and $2.25\units{s}$ for the fourth order method. These values were found empirically by increasing the step size in increments of $0.25\units{s}$ until the simulation was unable to successfully reach the final time of $1500\units{s}$. As discussed in Sec.~\ref{sec:methods}, the ImEx schemes are limited by the stability of the explicit portion of the method. In general, the higher order ImEx methods have a larger explicit linear stability region and can allow for larger time step sizes (at additional computational expense).

Figure~\ref{fig:time_integration_performance} gives the time-to-solution for simulations to reach the end of the two particle test case ($t=1500\units{s}$) versus the $L_2$ error (as defined for Fig.~\ref{fig:time_integration_convergence}) on one node of the OLCF Frontier supercomputer using all eight of the GPUs available. Time-to-solution comparisons between the manually implemented first-order method and the SUNDIALS implementation of the same method yielded run times that differed by less than the run-to-run variation observed ($\sim$7\%). Thus, we incur no appreciable overhead utilizing the SUNDIALS library and only report results from the more flexible SUNDIALS implementations for the remainder of this article. The work-precision diagram (Fig.~\ref{fig:time_integration_performance}) enables us to determine the most efficient approach for a particular level of accuracy. The level of error is a key criterion for scientific investigations using phase-field simulations; a faster time-to-solution may not be preferred if it causes an increase in error. However, this preference for lower error is not absolute; below a certain level of error, the changes in the solution cease to be useful for scientific insight or decision making. In Fig.~\ref{fig:time_integration_performance} the region below the physically useful error is shown in blue. Correspondingly, the data points in the white region in the figure may have time integration error that is too large for the simulation results to be useful. The physically useful error is inherently application-dependent and subjective. The criterion we choose for our analysis is whether the interface contour (i.e., the 0.5 level set of the order parameter field) in a slice down the center of the larger of the two particles is clearly distinguishable from the contour for the highly resolved reference solution when viewing the full extent of the particle. This criterion corresponds to convergence of a key quantity of interest: the shape of the simulated particles. Simulations using the first order method reach this criterion between the $\Delta t = 0.25\units{s}$ and $\Delta t = 0.125\units{s}$. Interpolating between these points, we place the upper bound on the physically useful error at $0.40$ for a first-order simulation with $\Delta t = 0.35\units{s}$.

For simulations using the tight elastic solver tolerance, as used for the temporal convergence tests, the time-to-solution increases approximately linearly in the log-log plot with decreasing error. For a similar time-to-solution the first order simulations have much higher error than the higher-order simulations. The higher order simulations are bunched in a band, with only minor variation in the time-to-solution for similar levels of error. Note that even the data points with the highest error for the second through fourth order methods are well below the physically meaningful level. While one may expect that a shorter time-to-solution could be achieved with these methods by using larger time step sizes to obtain solutions with larger errors still below the useful threshold, the maximum time step possible is limited by the stability of the explicit terms in the ImEx splitting. Since all second- through fourth-order simulations have adequate accuracy, the best performing simulation is simply the one with the shortest run time. Therefore, for the tight elastic solver tolerance, the best performing simulation is the third-order simulation with $\Delta t = 1.5\units{s}$ with a time-to-solution of $800\units{s}$. This time-to-solution is approximately $2.7\times$ faster than the only first-order result with error below the physically useful level ($\Delta t = 0.125\units{s}$ and a time-to-solution of $2192\units{s}$) and is 1.5$\times$ faster than the next first-order result ($\Delta t = 0.25 \units{s}$ and a time-to-solution of 1197$\units{s}$) which has an unacceptable level of error.

While these simulations with a tight elasticity solver tolerance are appropriate for calculating the temporal discretization error, a looser tolerance reduces the time-to-solution significantly without adding physically relevant error. Figure \ref{fig:time_integration_performance} also includes simulation results with a loose elasticity solver tolerance ($\epsilon_{el} = 10^{-6}$). With this tolerance, the elasticity solver exits after one iteration for nearly all time steps; further loosening of the tolerance does not lead to appreciable reduction in time-to-solution. With the change to the loose tolerance, the time-to-solution drops by about an order of magnitude with the overall range of times-to-solution going from between $646\units{s}$ and $1415\units{s}$ down to between $63\units{s}$ and $167\units{s}$. The additional error introduced by the elasticity solve has negligible effect on the first order results due to their already large error. For the higher order simulation results, the looser tolerance increases the $L_2$ error and compresses the range of errors, but the results are still well below the acceptable error. With the loose tolerance, the best performing simulation is the fourth order simulation with $\Delta t = 2.25\units{s}$ with a time-to-solution of 87 s. This time-to-solution is approximately $2.9\times$ faster than the only first-order result with error below the physically useful level ($\Delta t = 0.125\units{s}$ and a time-to-solution of $250\units{s}$) and is 1.4$\times$ faster than the next first-order result ($\Delta t = 0.25 \units{s}$ and a time-to-solution of 125 $\units{s}$) which has an unacceptable level of error.

In Fig.~\ref{fig:time_integration_performance}, we also present simulation results using the fourth-order integrator with adaptive time step sizes. The tolerances for adaptivity are applied pointwise to the change in the (unitless) order parameter field. Three values of the absolute tolerance were chosen: $2 \times 10^{-9}$, $5 \times 10^{-9}$, and $1 \times 10^{-8}$. 
A relative tolerance of $1 \times 10^{-9}$ was used for all tests, with the result being insensitive to its value. Of these simulations, the $5 \times 10^{-9}$ absolute tolerance value resulted in the fastest time-to-solution, $87\units{s}$, the same time-to-solution as the fastest fixed time step simulation, albeit with slightly reduced $L_2$ error. 
The simulations with adaptive time steps have the same stability limitations as the fixed time step simulations.  Here, the error increases rapidly above $\Delta t = 2.25\units{s}$ due to instability, and the time step size controller keeps the time step sizes at or below this level. Although adaptive time stepping does not improve time-to-solution, these findings point to a robustness benefit -- the adaptive scheme automatically finds the appropriate time step without manual trial-and-error with different fixed values of $\Delta t$.

The time-to-solution for these simulations is dominated by the number of right-hand side (RHS) evaluations. The total number of RHS evaluations is the product of the number of RHS evaluations per time step and the number of time steps. Because of the choice of ImEx splitting (see Sec.~\ref{sec:methods}) the explicit and implicit RHS functions are evaluated the same number of times each step as RHS evaluations are not necessary as part of the implicit stage solves. The number of RHS evaluations per time step is as follows: one for the first-order method, three for the second-order method, four for the third-order method, and six for the fourth-order method. Examining the results with the loose elasticity tolerance, the fourth-order method with the highest stable time step size (six RHS evaluations per time step, 667 time steps) has three times fewer total RHS evaluations than the first-order simulation at the upper bound of the physically relevant error (one RHS evaluation per time step, 12,000 time steps), almost exactly matching the observed $2.9\times$ speedup in time-to-solution. This relationship is somewhat weaker for the tight elastic tolerance, where the variable number of iterations in the nonlinear elasticity solve cause variation in the time per RHS evaluation (e.g., the speedup between the fourth-order, $\Delta t = 2.25\units{s}$ simulation compared to the first-order, $\Delta t = 0.125\units{s}$ simulation is only $2.7\times$).

In summary, using an ImEx-ARK time integration approach, we demonstrated expected convergence behavior for first- through fourth-order schemes and a $2.9\times$ speedup under scientifically relevant simulation conditions using a fourth-order scheme compared to the widely used first order scheme. The second- and third-order order schemes also provided a speedup relative to the first-order scheme. All stable simulations using the second- through fourth- order schemes exhibited $L_2$ error well below the physically relevant level. Therefore, in addition to reduced time-to-solution compared to the first-order simulations (on the edge of the physically relevant error), these results suggest that any simulation with the second- through fourth-order schemes that are stable, also have acceptable error. Given these findings, the fourth-order scheme was used for the other studies in this article.


\subsection{Limitations and future work} \label{sec:limitations_and_future_work}

While the speedup demonstrated in the previous section is already paradigm shifting, further improvements are likely possible. The time integrator study in Sec.~\ref{sec:time_integration} and the large-scale growth and coarsening simulation highlight that the higher order ImEx methods enable larger time step sizes and produce errors well below the physically relevant level. However, the maximum step size is still limited by stability not the solution error.
This suggests alternative partitionings of the right-hand side (e.g., treating some of the nonlinear terms implicitly) may improve stability and allow the use of larger time step sizes while still having error below the physically relevant level. The challenge with such an approach will be ensuring an efficient nonlinear solve for the implicit stages to ensure gains in step size are not offset by increased solver costs. 
Second, the fourth-order time integration scheme exhibited order-reduction as the time step size decreased. When tighter error control is required, developing ImEx methods with higher weak stage order (building on the work in \cite{biswas2025explicit, biswas2023design, biswas2024algebraic} to alleviate order reduction with explicit and implicit methods) may yield improved convergence.

The ability to take larger time steps near the physically relevant error level would increase the likelihood of performance gains from adaptive time integration. A preliminary test using adaptive time integration for the same simulation conditions in Fig. \ref{fig:demo} revealed similar results to what was seen in Sec. \ref{sec:time_integration} -- a slightly larger time step size ($\Delta t = 2.28$s vs. $2$s) but a slightly longer time-to-solution (5.6h vs. 5.5h) due to overheads associated with adaptive time integration.

Two improvements to node-level performance may be worth investigation. First, the performance of the Kokkos \texttt{MDRangePolicy} construct is known to be sensitive to tiling \cite{kokkos-docs-mdrange}. Autotuning the tiling parameter may result in increased performance for time spent in Kokkos loops. We did not investigate this type of optimization in this article because it necessarily depends on the grid size and computational resources -- instead our focus is on general acceleration strategies. Second, we have not implemented Kokkos host-parallel (e.g. OpenMP or C++ Threads) backends and have focused on the more performant GPU backends. Using shared-memory parallelism for CPU-only calculations may provide a speedup compared to our use of intranode MPI communication. However, we expect any speedup to be limited, on the order of the 6-16\% of the total time spent in MPI communication in Tables \ref{tab:summit-performance} and \ref{tab:frontier-performance}.

More detailed profiling of scaling tests may identify the portions of the code causing the current scaling limits. Identification of bottlenecks for weak scaling are particularly of interest due to the substantially sub-$\mathcal{O}(N^3 \log(N))$ scaling. However, at the application level it is unclear what optimizations could change the scaling behavior, since nearly all MPI communication is performed during FFT library calls. If the sub-ideal weak scaling is due to GPU register pressure, as hypothesized in Sec.~\ref{sec:scaling}, reducing the number of variables stored on the GPU may improve scaling performance (albeit potentially at the expense of recomputing previously stored fields).

With the large simulation domains now accessible with MEUMAPPS C++, visualization becomes a challenge. One of the reasons for performing the demonstration simulation in a 1152$^3$ domain rather than a 2304$^3$ was due to our estimate that rendering 70 frames for visualizing would take days of computing time with the standard HPC visualization tool VisIt \cite{HPV:VisIt}. In-situ visualization (where visualization calculations are performed during a simulation with co-allocated resources), may provide a tractable path forward for large simulations with MEUMAPPS C++ \cite{10767626}.

Although this article focuses on the implementation of the phase field model for solid-state phase transformation developed by Radhakrishnan et al.~\cite{Radhakrishnan2016}, MEUMAPPS C++ is designed in a modular fashion to allow rapid development of different physics applications. One straightforward extension of the current model is the incorporation of the more complex CALPHAD representation of the homogeneous free energy functions (e.g., using the Thermo4PFM library \cite{FATTEBERT2023108739}) rather than the parabolic functions used here. Phase-field models for phenomena other than precipitation can also be implemented with MEUMAPPS C++. For example, the MEUMAPPS C++ repository \cite{doecode_69332} contains an application to solve the Cahn-Hillard equation for the PFHub benchmark problem \cite{JOKISAARI2017139}. Pseudospectral methods are extensively used for phase-field crystal models \cite{Emmerich01122012}, making them an attractive set of models to implement with MEUMAPPS C++. Beyond phase-field models, other models often solved using pseudospectral methods such as computational fluid dynamics and crystal plasticity are also candidates for implementation with MEUMAPPS C++.

\section{Conclusion}

In this article we demonstrated a holistic approach to accelerate pseudospectral phase-field simulations using a combination of  GPU acceleration, distributed memory scaling, and high-order time integration implemented in the MEUMAPPS C++ framework. Substantial GPU acceleration was shown on both NVIDIA and AMD GPUs with a 4.7-9.3 $\times$ speedup observed on NVIDIA GPUs (comparing 6 NVIDIA V100 GPUs to 42 CPU cores) and a 7.6-17.5 $\times$ speedup observed on AMD GPUs (comparing 8 AMD MI250X GPUs to 64 CPU cores). Using performance portability layers in MEUMAPPS C++, these calculations on CPUs and both types of GPUs required no hardware-specific code. Scaling tests showed near-ideal strong scaling up to 4096 GPUs for a 2304$^3$ domain and effective use of hundreds of GPUs for smaller domains. Switching from the first-order implicit-explicit scheme typically used to a fourth-order scheme leads to a 2.6-2.9 $\times$ speedup, with error reduced to well below the physically meaningful level. We demonstrated that this holistic approach drastically accelerates an application-relevant simulation that investigates the growth and coarsening of 1920 $\gamma''$ precipitates in a Ni-Nb-Fe alloy. A simulation of a 7h heat treatment of this system was completed in just 5.5 hours. We presented an estimate that this is a 217-501$\times$ speedup over a baseline approach (a CPU-only calculation using first order time integration running on 512 cores). Therefore, a calculation that would normally take months can be completed in an afternoon.

While high-order time integration, GPU acceleration, and distributed memory scaling individually are not new, we demonstrate how recent advances in these areas lead to transformative performance improvements when combined. Furthermore this demonstration was not for a toy problem; by implementing these approaches in MEUMAPPS C++ we tested these approaches for a full-physics application-relevant simulation of alloy heat treatment. The approach presented in this article makes 3D simulations of thousands of precipitates tractable. Such large-scale simulations are necessary to quantitatively evaluate coarsening behavior, examine long-range interactions (e.g., rafting), to examine the collective behavior of systems with many phases and/or orientation variants, and many other phenomena. Even for smaller-scale problems, the performance improvements presented here enable more advanced uses of phase-field simulations, such as large ensembles of simulations to quantify uncertainty and integration of faster-than-real-time simulations into experiments. The holistic approach to accelerating pseudospectral calculations and the MEUMAPPS C++ implementation are relevant to a wide range of applications, not only the phase-field simulations of precipitation that are the focus of this article. Other uses of phase-field models (e.g., grain evolution and phase-field crystal), fluid dynamics, and crystal plasticity are particularly relevant uses. Across all of these applications, the ability to leverage hundreds or thousands of GPUs will become increasingly important as AI applications motivate the continued proliferation of GPU-accelerated computing computing clusters. The combination of approaches for GPU acceleration, parallel scaling, and time integration presented here give researchers using pseudospectral methods the capability to take full advantage of these computational resources and thereby unlock new insights with their simulations.

\section{Methods} \label{sec:methods}
\subsection{Phase-field model formulation}
In this article we use a Kim-Kim-Suzuki (KKS) phase-field model of solid-state phase transformations for our test cases. Specifically, we use the multi-component, multiphase MEUMAPPS-SS model described in Refs. \cite{PhysRevMaterials.5.053401, Radhakrishnan2016}. We refer readers to those references (and the references therein) for a detailed description of the model; here we briefly lay out the model equations. The total free energy is given by
\begin{equation}
    \mathcal{F} = \int (f_b + f_{el} + f_{ch})~d\mathcal{V},
\end{equation}
where each component of the free energy density is given by
\begin{equation}
    f_b = \frac{1}{2} \sum_{v} \left[\nabla \phi_v \right]^T \kappa_v \left[\nabla \phi_v \right] 
+ \sum_{v} \sum_{q>v} \omega_{vq} |\phi_v \phi_q| 
+ \bar{\omega} \sum_{q} \phi_q \left(1 - \sum_{v} \phi_v \right),\label{eq:f_b}
\end{equation}
\begin{equation}
    f_{el} = \sum_{v} h(\phi_v) f_{el}^v + \left[1-\sum_{v} h(\phi_v) \right] f_{el}^{matrix},
\end{equation}
\begin{equation}
    f_{ch} = \sum_{v} h(\phi_v) f_{ch}^v + \left[1-\sum_{v} h(\phi_v) \right] f_{ch}^{matrix} \label{eq:f_ch},
\end{equation}
where
\begin{equation}
    h(\phi) = \phi^3 (6\phi^2 - 15\phi + 10),
\end{equation}
\begin{equation}
        f^{p}_{el} = \frac{1}{2} \epsilon_{ij}^{el,p} : C_{ijkl}^p : \epsilon_{kl}^{el,p},
\end{equation}
\begin{equation}
        \epsilon_{ij}^{el, p} = \bar{\epsilon}_{ij} + \frac{1}{2}\left(\frac{\partial u_i}{\partial x_j} + \frac{\partial u_j}{\partial x_i} \right) - \epsilon_{ij}^{*, p}, \label{eq:epsilon_el}
\end{equation}
\begin{equation}
    f_{ch}^p = A^p + \sum_c \left[ A^{p,c} (X^{p,c} - B^{p,c}) \right]. \label{eq:f_ch_p}
\end{equation}
The index $v$ refers to precipitate phases and the index $p$ refers to all phases (i.e., the precipitate phase and the matrix phase). The index $c$ refers to solute elements. Note, we made slight changes in notation (especially in equation \eqref{eq:f_ch}) from \cite{PhysRevMaterials.5.053401, Radhakrishnan2016} for a general number of phases and components. For equations \eqref{eq:f_b} to \eqref{eq:epsilon_el}, the following model constants were introduced: $\kappa_v$ (gradient energy tensor per phase), $\omega_{vq}$ (pairwise barrier), $\bar{\omega}$ (double-well barrier), $A^p$, (homogeneous free energy offset), $A^{p,c}$ (homogeneous free energy curvature), $B^{p,c}$ (homogeneous free energy reference composition), $C^p_{ijkl}$ (single-phase stiffness tensor), $\bar{\epsilon}$ (mean strain), and $\epsilon_{ij}^{*, p}$ (single-phase eigenstrain).

The time evolution equations are derived using the standard variational approach for a KKS model, assuming conservation of the composition fields and no conservation of the order parameter fields. For the composition fields, $X^c$, this yields
\begin{equation}
    \frac{\partial X^c}{\partial t} = \nabla \cdot \left ( M^c \nabla \frac{\partial f_{ch}}{\partial X^c} \right), \label{eq:x_gov}
\end{equation}
where
\begin{equation}
    X^c = \sum_{v} h(\phi_v) X^{v,c} + \left[1-\sum_{v} h(\phi_v) \right] X^{matrix,c}, \label{eq:x_sum}
\end{equation}
and $M^c$ is the mobility for element $c$. As is standard in KKS models, the $X^{p,c}$ are constrained such that the chemical potentials of an element are equal in all phases at the interface,
\begin{equation}
    \frac{\partial f^{p}}{\partial X^{p,c}} = \frac{\partial f^{p'}}{\partial X^{p',c}}. \label{eq:equal_mu}
\end{equation}

For the time evolution of the order parameter fields, $\phi_v$, we make a simplifying assumption from Refs. \cite{PhysRevMaterials.5.053401, Radhakrishnan2016}. In those references the MEUMAPPS-SS model is derived in a multi-phase-field context, where $\sum_p \phi_p = 1$, which results in an expression that depends on the pointwise number of non-zero fields \cite{SHI20136006}. Such a formulation is prone to numerical artifacts as $\phi_v$ values cross the threshold for being ``non-zero''. Therefore, in this work, we use the simpler multi-order-parameter expression, which has no assumption on $\sum_p \phi_p$ \cite{DEWITT2017378, JI2014259}, given by
\begin{equation}
    \frac{\partial \phi_v}{\partial t} = - L \frac{\delta \mathcal{F}}{\delta \phi_v}, \label{eq:phi_gov}
\end{equation}
where $L$ is the interface mobility.
The previous use of the multi-phase-field expression was motivated by the desire to simulate overlapping nuclei for variants in Ti-6Al-4V to examine competition between orientation variants \cite{Radhakrishnan2016} and the $\sum_p \phi_p = 1$ constraint was important to maintain sensical interpolated values in these overlap regions. For the test cases in this article, with the more typical situation of well-separated precipitates that never overlap, the multi-order-parameter expression is sufficient (and should lead to near-identical results as the multi-phase-field expression).

The final governing equation for the displacement field, which is assumed to be a quasi-equilibrium at all times \cite{Durga_2013}, is given by
\begin{equation}
    \nabla \cdot \left\{ C_{ijkl} \left[ \bar{\epsilon}_{ij} + \frac{1}{2}\left(\frac{\partial u_i}{\partial x_j} + \frac{\partial u_j}{\partial x_i} \right) - \left( \sum_{v} h(\phi_v) \epsilon_{ij}^{*,p} \right) \right] \right\} = 0, \label{eq:u_gov}
\end{equation}
where 
\begin{equation}
    C_{ijkl} = \left\{ \sum_{v} h(\phi_v) \left[C_{ijkl}^v \right]^{-1} + \left[ 1-\sum_{v} h(\phi_v) \right] \left[C_{ijkl}^{matrix} \right]^{-1} \right\}.
\end{equation}
In summary, the governing equations for the MEUMAPPS-SS model used in this article are given by \eqref{eq:x_gov}, \eqref{eq:phi_gov}, and \eqref{eq:u_gov}.

\subsection{MEUMAPPS C++ and the MEUMAPPS-SS application}
In this article, we implement the model described in the preceding section in the MEUMAPPS-SS application using the MEUMAPPS C++ framework for solving systems of PDEs with the pseudospectral method and the numerical and computational approaches laid out in this article. A preliminary version of MEUMAPPS C++ was described in Ref.~\cite{doi:10.1177/10943420211042558}. MEUMAPPS C++ and its MEUMAPPS-SS application generalize and extend the original MEUMAPPS-SS code \cite{doecode_45884, Radhakrishnan2016}, which was written in Fortran and was used in Refs.~\cite{Radhakrishnan2016, osti_1883850, PhysRevMaterials.5.053401}. MEUMAPPS C++ was restructured to be similiar to the core library/application structure of the PRISMS-PF framework \cite{prismspf}, where the core library contains shared functionality (e.g., library interfaces, generic input/output, initial condition utilities, and a mechanics solver) and the applications contain the implementation of functionality tied to a particular use case (e.g., use-case-dependent input parameters, governing equations, and postprocessing analysis). In addition to the core library and the MEUMAPPS-SS application, the MEUMAPPS C++ repository \cite{doecode_69332} contains applications for the Allen-Cahn equation, the Cahn-Hilliard equation (PFHub Benchmark 1a \cite{JOKISAARI2017139}), and a simple coupled Allen-Cahn-Cahn-Hilliard system. Although application development for MEUMAPPS C++ has been focused on phase-field models, the framework is designed to be general for other models amenable to pseudospectral methods.

Next, we discuss briefly the general numerical approach used to solve the governing equations, with a more detailed description of some aspects in later sections. We apply a pseudospectral discretization in space, using FFTs to move from real space (where the physical inputs and outputs are defined) to reciprocal space (where the time integration is performed). In reciprocal space the PDE solve over the entire domain is replaced with ODEs at every grid point. Following a common approach in pseudospectral phase-field simulations since the seminal work by Chen and Shen \cite{CHEN1998147}, we use an ImEx scheme to advance the ODE system in time. This choice of time integration approach allows us to exploit the structure of \eqref{eq:x_gov} and \eqref{eq:phi_gov} for greater efficiency. The reciprocal space counterparts of \eqref{eq:x_gov} and \eqref{eq:phi_gov} can be rewritten as
\begin{equation} \label{eq:composition}
    \frac{\partial \tilde{X^c}}{\partial t} = M^c |\mathbf{k}|^2 \tilde{X^c} + F_{nonlinear}^X,
\end{equation}
\begin{equation} \label{eq:order}
    \frac{\partial \tilde{\phi^v}}{\partial t} = L \left(\sum_{i,j} \kappa_{i,j} k_i k_j \right)\tilde{\phi^v} + F_{nonlinear}^\phi,
\end{equation}
where the right-hand side is partitioned into a stiff diffusive term and a less stiff nonlinear term. The diffusive term is linear and treated implicitly, while the nonlinear terms are handled explicitly. With the pseudospectral discretization, the linear solve necessary for the implicit portion of the time integration method is trivial to compute so the time advance can proceed without any (nontrivial) linear or nonlinear solves.

For a general form of the single-phase homogeneous free energy density, $f_{ch}^p$, satisfying Eq.~\eqref{eq:x_sum} and \eqref{eq:equal_mu} requires the solution of a pointwise nonlinear equation \cite{PhysRevE.60.7186}. However, following a common choice in the literature \cite{PhysRevMaterials.5.053401, PhysRevE.60.7186, Radhakrishnan2016, DEWITT2017378, JI2014259} we use parabolic single-phase homogeneous free energy densities (see Eq.~\eqref{eq:f_ch_p}), such that the single phase compositions are given by an analytical expression. To solve Eq.~\eqref{eq:u_gov}, nonlinear mechanical equilibrium, we use the iterative approach of Hu and Chen \cite{HU20011879, Durga_2013}.

\subsection{Performance portability layer} \label{sec:kokkos}

One of the contributions of this work is to demonstrate the benefit of performance portability approaches to effectively utilize diverse modern computing hardware. Performance portability is the concept of developing software that runs on a variety of computing hardware without duplicate work and doing so without significant sacrifices in performance versus software specifically targeting the hardware \cite{9502936}. Performance portability focuses on on-node parallelism (rather than between-node distributed memory parallelism which is typically accomplished with a Message Passing Interface (MPI) library). The current increase in diversity of architectures for supercomputers highlights the need for performance portability; as of writing the top ten largest supercomputers include systems with NVIDIA/CUDA, AMD/HIP, and Intel/oneAPI hardware \cite{top500_nov2025}.

In this work we use the Kokkos library \cite{9502936, 9485033} as our performance portability layer. Kokkos provides MEUMAPPS C++ with data structures and parallel execution operations that have architecture-specific backends. The specific data layouts and execution patterns used depend on the backend but come from the same Kokkos calls in MEUMAPPS C++. Thus, very little of MEUMAPPS C++ has architecture-specific code, and the architecture-specific code it has is largely limited to precompiler macros that provide the proper header files to include. MEUMAPPS C++ has support for the serial, CUDA, and HIP backends in Kokkos (targeting CPU, NVIDIA GPU, and AMD GPU architectures, respectively). Support for the experimental SYCL backend (for Intel GPUs) is planned and is not expected to require significant development. Support for the experimental OpenMP target backend (using OpenMP for GPU offload) is not planned, with architecture-specific backends being preferred, but adding support is not expected to require significant development. MEUMAPPS C++ does not currently support the host parallel backends (OpenMP, C++ Threads, and HPX), as the focus is on GPU utilization. MEUMAPPS C++ relies heavily on Kokkos data structures, with the \texttt{Kokkos::View}, a type of multidimensional array, being the data structure used for storing all field variables. As shown in Tables \ref{tab:summit-performance} and \ref{tab:frontier-performance}, kernels written as Kokkos lambda functions or functors occupy the vast majority of the run-time of the code in MEUMAPPS C++ itself (i.e., not counting the library calls to heFFTe for the distributed 3D FFTs). 

One of the contributions in this work is an approach for translating between \texttt{View}s and the data structures for the other libraries we use, SUNDIALS and heFFTe. The translation approach is discussed in the following two subsections.

\subsection{Time integration} \label{sec:sundials}
To advance the model equations in time, we utilize ImEx-ARK time integration methods from the ARKODE \cite{reynolds2023arkode} package in the SUNDIALS \cite{gardner2022enabling,hindmarsh2005sundials} library of time integrators and nonlinear solvers. 
Unlike the multi-step method presented in Ref.~\cite{CHEN1998147}, ImEx-ARK methods are one-step, multi-stage methods. 
Our primary motivation for choosing ImEx-ARK methods is their greater stability at high order \cite{rosales2017unconditional,frank1997stability}, the potentially lower cost of error estimates for step size adaptivity with lower order methods compared to step-doubling approaches with ImEx multi-step methods \cite{mara2025performance}, and the availability of an established, general-purpose, GPU-accelerated library (ARKODE) for C++ applications.

ImEx-ARK methods target mixed stiff/non-stiff systems of the form
\begin{equation} \label{eq:imex_ode}
    y' = f^{I}(t, y)+ f^{E}(t, y), \quad y(t_0) = y_0,
\end{equation}
where $y$ is the state vector of composition fields and order parameters with the initial condition $y_0$ at time $t_0$. 
The right-hand side function $f^{I}$ contains the stiff linear terms in Eqs.~\eqref{eq:composition} and \eqref{eq:order} which are treated implicitly while $f^{E}$ includes the remaining nonlinear terms which are treated explicitly.
A step from $t_{n-1}$ to $t_{n}$ with an ImEx-ARK method is given by
\begin{subequations}
  \label{eq:ARK_method}
  \begin{align}
    \label{eq:ARK_stage}
    z_i &= y_{n-1} + \Delta t_n \sum_{j=1}^{i} A^I_{i,j} f^I(t^I_{n,j}, z_j) + \Delta t_n \sum_{j=1}^{i-1} A^E_{i,j} f^E(t^E_{n,j}, z_j), \quad i=1,\ldots,s, \\
    y_n &= y_{n-1} + \Delta t_n \sum_{i=1}^{s} \left( b^I_i f^I(t^I_{n,i}, z_i) + b^E_i f^E(t^E_{n,i}, z_i) \right),
  \end{align}
\end{subequations}
where the method for computing the internal stages, $z_i$, and the new solution, $y_n$, is defined by a pair of $s$-stage Butcher tables containing the implicit ($A^I \in \mathbb{R}^{s \times s}$, $b^I \in \mathbb{R}^s$, $c^I \in \mathbb{R}^s$) and explicit ($A^E \in \mathbb{R}^{s \times s}$, $b^E \in \mathbb{R}^s$, $c^E \in \mathbb{R}^s$) coefficients.
The implicit and explicit internal stage times are denoted by $t^I_{n,j} = t_{n-1} + c^I_j \Delta t_n$ and $t^E_{n,j} = t_{n-1} + c^E_j \Delta t_n$, respectively, with step size $\Delta t_n = t_{n} - t_{n-1}$.
In general, computing $z_i$ requires solving a nonlinear algebraic system of equations.
However, by treating just the stiff, linear terms implicitly we only need to solve a linear system for each stage.
Moreover, the pseudospectral discretization makes the resulting linear solves trivial to compute.

For some methods, the Butcher table may include an additional set of coefficients, $\tilde{b}^I \in \mathbb{R}^s$ and $\tilde{b}^E \in \mathbb{R}^s$, that define a (typically) lower-order embedded solution,
\begin{equation} \label{eq:ARK_embedding}
    \tilde{y}_n = y_{n-1} + h_n \sum_{i=1}^{s} \left( \tilde{b}^I_i f^I(t^I_{n,i}, z_i) + \tilde{b}^E_i f^E(t^E_{n,i}, z_i) \right).
\end{equation}
The difference between $y_n$ and $\tilde{y}_n$ provides an estimate of the local truncation error (LTE) in a step which can be utilized to adapt the step size. An attempted step is accepted if $\|{\rm{LTE}}\|_{\textsc{wrms}} \leq 1$ in the weighted root-mean-square (WRMS) norm,
\begin{equation} \label{eq:wrms_norm}
    \| v \|_{\textsc{wrms}} = \left( \frac{1}{N} \sum_{i=1}^{N} \left( v_i w_i \right)^2 \right)^{1/2},
\end{equation}
where $N$ is the length of the vector $v$ and the weights $w_i$ are given by 
\begin{equation} \label{eq:wrms_weights}
w_i =\big(\mathrm{rtol}\, \lvert y_{n-1,i} \rvert +\mathrm{atol} \big)^{-1},
\end{equation}
with the relative and absolute tolerances, rtol and atol, respectively. If a step attempt is rejected, a new step size is computed based on the LTE, and the step is repeated. After successfully completing a step, the LTE is similarly used to determine the step size for the next step attempt. The results presented above utilize the first order ImEx Euler method, the second order method from Ref. \cite{giraldo2013implicit}, as well as the third and fourth order methods from Ref. \cite{kennedy2003additive}. The second through fourth order methods include embeddings for adaptive step size selection.

Applying an ImEx splitting to Eqs.~\eqref{eq:composition} and \eqref{eq:order} enables larger time steps than possible with a fully explicit method as treating the stiff linear terms implicitly alleviates the primary stability restriction on the step size. 
This choice of splitting also avoids the complexity and expense of a nonlinear solve in each stage when an implicit method is applied to all of the terms. As such, the trade-off with this choice of splitting is that implicit solves can be computed quickly and easily, but the time step sizes are still subject to the stability limitations of the nonlinear terms.

Interfacing with SUNDIALS requires, at a minimum, providing the time integrator with a vector object to operate on state data, $y$, and functions to evaluate $f^I$ and $f^E$. 
As noted above, the problem state is stored in separate multidimensional Kokkos \texttt{View}s for $X^c$ and $\phi^v$. 
The local data underlying each \texttt{View} is a one-dimensional array, passed with a pointer, that is trivially wrapped by the native SUNDIALS vector objects (\texttt{NVectors}) which support NVIDIA, AMD, and Intel GPUs \cite{balos2021enabling}. The SUNDIALS update is indifferent to the ordering of elements in the one-dimensional array because the time integration is a pure point-wise operation.
The distributed state vector (i.e., $y$) is formed by constructing an MPI-aware \texttt{ManyVector} \cite{reynolds2019sundials} (itself an \texttt{NVector}) comprised of the local subvectors wrapping the \texttt{View} data.
When evaluating $f^I$ or $f^E$, the state data wrapped by the input \texttt{NVector} is copied into the MEUMAPPS-SS objects for computing the FFTs and evaluating the relevant physics. 
Relative to the other computations in MEUMAPPS-SS, this data copy incurs no meaningful overhead.
Finally, to take advantage of the special structure of the linear solves with a pseudospectral discretization, we wrap the MEUMAPPS-SS linear solve computation as a SUNDIALS linear solver object (\texttt{SUNLinearSolver}) enabling ARKODE to directly call the MEUMAPPS-SS native linear solver for the Poisson systems in each implicit stage.

\subsection{Scalable, performance-portable fast Fourier transforms} \label{sec:heffte}

To compute multidimensional distributed FFTs, MEUMAPPS C++ relies on the highly efficient fast Fourier transforms for exascale (heFFTe) library \cite{10.1007/978-3-030-50371-0_19} which unifies multiple parallel FFT approaches found in different DOE applications \cite{ayala2021scalability, ayala2022fft}. 
The two primary kernels underlying parallel FFTs are (1) computing one-dimensional transformations and (2) data reshaping operations to align the data such that all the entries along a single dimension belong to the same MPI rank. 
To compute the one-dimensional FFTs, heFFTe interfaces with a number of different on-node backends such as FFTW \cite{FFTW05} and vendor optimized libraries for AMD, Intel, and NVIDIA GPUs. 
As such, the main focus of heFFTe is on the intricacies of the communication algorithms for data reshaping operations which usually exceed the cost of the one-dimensional transformations, especially with GPU acceleration \cite{ayala2022performance, ayala2022analysis}.


The heFFTe interface operates on flat arrays of local data with a corresponding \texttt{box3d} object that describes the global indices of the data and the data ordering. As the data underlying the Kokkos \texttt{View}s are one-dimensional arrays, it is straightforward to pack/unpack the state data into a contiguous array for FFT computations. Unlike with the SUNDIALS integration, the heFFTe integration requires specific ordering of the contiguous array to maintain the spatial relationships in each field. For consistency, arrays in MEUMAPPS C++ are always passed to heFFTe using the Kokkos \texttt{LayoutLeft} ordering. The fields with FFTs applied to them in MEUMAPPS C++ are usually four- or five-dimensional Kokkos \texttt{Views}, with related fields (e.g., the vector of composition fields) being stacked into a single \texttt{View}. Each three-dimensional field is individually deep copied from the higher dimensional \texttt{View} into a temporary \texttt{LayoutLeft} \texttt{View} to pass into heFFTe. For \texttt{View}s residing on the GPU, the default layout is \texttt{LayoutLeft} whereas it is \texttt{LayoutRight} when residing on the CPU. With \texttt{View}s declared with the proper layout, the \texttt{Kokkos::deep\_copy} operation automatically changes the layout during the deep copy. After the FFT operation is completed, the field is deep copied back into the appropriate higher dimensional \texttt{View}. In principle, the deep copy is not necessary and a pointer to the appropriate memory location inside the higher dimensional \texttt{View} could be passed to heFFTe along with whether it is ordered \texttt{LayoutLeft} or \texttt{LayoutRight}. However, the structure of the higher-dimensional views complicates the striding of the array data in memory across the different fields, and so further examination of removing these deep copies is left to future work. Profiling indicates that these deep copies are responsible for less than 5\% of the total run time across different domain sizes and hardware, and therefore do not substantially change the overall performance of the simulations.

heFFTe has several configuration options involving communication patterns and indexing during reshaping, some of which have a significant effect on performance. heFFTe supports ``slab'', ``pencil'', and ``brick'' domain decompositions. In this study, a ``pencil'' MPI decomposition was used for all variables (each MPI subdomain extended the full length of the domain along exactly one dimension). The default MEUMAPPS C++ algorithm was used to create the pencils based on the domain size and the number of processes, selecting the number of subdomains in the real-space y and z directions as the closest factor pair for the number of MPI ranks that results in equally sized subdomains (using the lower of the factors for y and the higher of the factors for z). To eliminate a reshape operation in the 3D FFT, the subdomain decomposition is rotated in reciprocal space, using the lower of the factors for x and the higher of the factors for y. This default decomposition strategy was used for all calculations in this article. heFFTe includes several communication patterns, of which pipelined point-to-point (``p2p\_pl'') and all-to-all with variably message sizes (``a2av'') were used in this article. Following standard practice, p2p\_pl was used for the single-node tests and a2av was used for the scaling tests and the large growth and coarsening simulation, although testing showed minimal differences in performance for the two strategies. heFFTe also includes an option to reorder the data for the 1D FFTs for contiguous memory layout (versus leaving it as-is and striding across the memory). Initial testing showed a slight performance benefit for using the reorder option, and thus that option was used for all simulations in this article.

The MEUMAPPS C++ implementation has a modular interface to FFT libraries, allowing straightforward extensibility to alternative libraries. In addition to heFFTe, MEUMAPPS C++ has experimental support for the AccFFT library \cite{gholami2015accfft}. However, initial testing revealed substantially improved performance for heFFTe over AccFFT and therefore only heFFTe results are cited in this article.

\subsection{Profiling}

To collect profiling data MEUMAPPS C++ utilizes Caliper \cite{Boehme_Caliper_Performance_Introspection_2016}, a performance analysis library that provides a general abstraction layer for code instrumentation and performance measurement in HPC applications. 
At the user level, Caliper defines a simple, low-overhead API for source-code annotation to mark regions of interest.
Behind this interface, Caliper can combine performance measurements from different sources with the annotation context to perform a wide range of performance analysis tasks and generate reports in human or machine-readable formats.
The report configuration can be set at runtime with built-in options for common workflows e.g., the ``runtime-report'' configuration collects timings for annotated regions. Some libraries commonly used in scientific computing, including Kokkos, have built-in support for Caliper.
The results presented in this article use the configuration options \texttt{runtime-report(profile.mpi,mem.highwatermark,profile.kokkos), calc.inclusive, mpi-report} to collect timings. In addition to the timings created by default, key regions of MEUMAPPS C++ were instrumented with custom marked regions. The timing information in Tables \ref{tab:summit-performance} and \ref{tab:frontier-performance} and Fig. \ref{fig:performance-detail} was generated from scripts analyzing Caliper text file outputs (\texttt{runtime-report(profile.mpi)} for MPI timing, \texttt{runtime-report(profile.kokkos)}, and custom marked regions for FFT timing).

\subsection{Computing systems} \label{sec:computing_systems}

The calculations in this article are performed on two supercomputers at the Oak Ridge Leadership Computing Facility (OLCF), Summit and Frontier. The Summit supercomputer contains 4608 IBM Power System AC922 nodes \cite{10.1007/978-3-031-32041-5_10, olcf_summit_user_guide}. Each node contains two IBM Power9 CPUs (totaling 42 usable cores, with 2 cores reserved for the operating system) and six NVIDIA V100 GPUs. Each Power9 CPU has 256 GB of memory (512 GB per node) and each V100 GPU has 16 GB of memory (96 GB per node). Each node has approximately 42 TFLOPS of theoretical throughput for double precision datatypes. The nodes are split into subdomains containing one CPU and three GPUs. Resources within a subdomain are connected with NVLink, and the subdomains are connected between the CPUs with an X-Bus. Summit nodes are connected using a dual-rail EDR InfiniBand network with a non-blocking fat-tree topology. The networking interface card for each node is connected to the CPUs.

The Frontier supercomputer contains 9856 HPE/Cray EX nodes \cite{10.1007/978-3-031-32041-5_10, olcf_frontier_user_guide}. Each node contains one 64-core AMD EPYC CPU and four AMD MI250X accelerators. Each MI250X accelerator contains two graphics compute dies; for practical purposes each compute die functions as a separate GPU. Following standard convention \cite{olcf_frontier_user_guide}, we refer to each MI250X graphics compute die as a GPU. The CPU on a Frontier node has 512 GB of memory and each GPU has 64 GB of memory (512 GB per node). Each node has approximately 191 TFLOPS of theoretical throughput for double precision datatypes.  All GPUs on the node are directly connected to each other and to the CPU. Frontier nodes are connected using a HPC Slingshot network with a dragonfly topology. Each Frontier node has four networking interface cards, connected to pairs of GPUs.

\section*{Author Contributions} \label{sec:contributions}
SD: Conceptualization, Methodology, Software, Validation, Formal analysis, Investigation, Writing - Original Draft, Writing - Review \& Editing, Visualization; DG: Conceptualization, Methodology, Software, Validation, Formal analysis, Writing - Original Draft, Writing - Review \& Editing; PF: Conceptualization, Methodology, Software, Writing - Review \& Editing; YS: Methodology, Software, Writing - Review \& Editing; MS: Methodology, Software, Writing - Original Draft, Writing - Review \& Editing, Funding acquisition; CW: Conceptualization, Methodology, Writing - Review \& Editing, Funding acquisition; BR: Conceptualization, Methodology, Software, Validation, Writing - Original Draft, Writing - Review \& Editing, Funding acquisition.

All authors read and approved the final manuscript.

\section*{Acknowledgements} \label{sec:acknowledgments}
This research was supported by the Exascale Computing Project (17-SC-20-SC), a collaborative effort of the U.S. Department of Energy Office of Science and the National Nuclear Security Administration. This research used resources of the Oak Ridge Leadership Computing Facility at the Oak Ridge National Laboratory, which is supported by the Office of Science of the U.S. Department of Energy under Contract No. DE-AC05-00OR22725. 
Support for this work was also provided in part by the U.S. Department of Energy, Office of Science, Office of Advanced Scientific Computing Research, Scientific Discovery through Advanced Computing (SciDAC) Program through the Frameworks, Algorithms, and Scalable Technologies for Mathematics (FASTMath) Institute.
This work was performed in part under the auspices of the U.S. Department of Energy by Lawrence Livermore National Laboratory under contract DE-AC52-07NA27344. LLNL-JRNL-2020868.
The authors also acknowledge guidance and advice from the Kokkos team, particularly Bruno Turcksin, Damien Lebrun-Grandi\'e, and Christian Trott.

\section*{Competing Interests} \label{sec:coi}
All authors declare no financial or non-financial competing interests. 

\section*{Data Availability} \label{sec:data}
The datasets generated and analyzed during the current study are available in the OLCF Constellation dataset repository, 10.13139/OLCF/3989587.

\section*{Code Availability} \label{sec:code}
The underlying code for this study is available in the Oak Ridge National Laboratory code repository and can be accessed via this link https://code.ornl.gov/meumapps/meumapps.

\bibliographystyle{ieeetr}
\bibliography{references}
\end{document}